\documentclass[epj]{webofc}
\usepackage[varg]{txfonts}   
\usepackage{epsf,epsfig,float,latexsym,rotating,graphicx,slashed,bm,ulem}
\usepackage{amsmath,amssymb,amsfonts}
\usepackage[dvipsnames,usenames]{color}
\usepackage{array,multirow,booktabs}
\newcommand{\beq}{\begin{equation}}
\newcommand{\eeq}[1]{\label{#1}\end{equation}}
\newcommand{\beqa}{\begin{eqnarray}}
\newcommand{\eeqa}[1]{\label{#1}\end{eqnarray}}
\newcommand{\eeqan}{\end{eqnarray}}

\newcommand{\CSnlo}{$\rm P_{\rm NLO}$}

\newcommand{\GOi}{$\rm M_{\rm I}$}
\newcommand{\GOii}{$\rm M_{\rm II}$}

\newcommand{\IHWnlo}{$\rm KM_{\rm NLO}$}
\newcommand{\MMii}{$\rm B_{2}$}
\newcommand{\MMiv}{$\rm B_{4}$}
\woctitle{MESON2016 - the 14$^\textrm{th}$ International Workshop on Meson Production, Properties and Interaction}
\begin{document}
\selectlanguage{english}
\title{Theoretical approaches to low energy $\bar{K}N$ interactions}

\author{Ale\v{s}~Ciepl\'{y}\inst{1}\fnsep\thanks{\email{cieply@ujf.cas.cz}} \and
        Maxim Mai\inst{2}  
}

\institute{Nuclear Physics Institute, 250 68 \v{R}e\v{z}, Czechia
\and
          The George Washington University, 725 21$^{\rm st}$ St. NW, Washington, DC 20052, USA
          }

\abstract{
We provide a direct comparison of modern theoretical approaches based 
on the SU(3) chiral dynamics and describing the low energy $\bar{K}N$ data. The model predictions 
for the $\bar{K}N$ amplitudes and pole content of the models are discussed.
}
\maketitle
%
\section{Introduction}
\label{sec:intro}

In our contribution we review the current status of low energy $\bar{K}N$ interactions and concentrate
on comparison of the available theoretical approaches derived from the effective SU(3) chiral Lagrangian 
that describes the interaction of the pseudoscalar meson octet with the ground state baryon octet. 
In the $S=-1$ sector the involved meson-baryon coupled channels are 
$\pi\Lambda$, $\pi\Sigma$, $\bar{K}N$, $\eta\Lambda$, $\eta\Sigma$ and $K\Xi$ 
with threshold energies from about $1250$ to $1810$ MeV. Since the $\Lambda(1405)$ resonance 
lies closely below the $\bar{K}N$ threshold the chiral perturbation series does not converge 
in its vicinity and coupled-channel re-summation techniques are standardly employed to sum 
the major part of the chiral expansion to obtain the scattering amplitude. 

In our recent work \cite{Cieply:2016jby} we performed a direct comparison of the theoretical approaches 
that include NLO corrections to the leading order in the chiral expansion and fix the free model parameters, 
the low energy constants, to reproduce the experimental data on $K^{-}p$ scattering and reactions including 
the recent precise measurement of kaonic hydrogen characteristics (the shift and width of the $1s$ level 
due to strong interaction) by the SIDDHARTA collaboration \cite{Bazzi:2011zj}. The discussed models comprise 
of 
\begin{itemize}
\item the Kyoto-Munich \cite{Ikeda:2012au} and Murcia \cite{Guo:2012vv} approaches, which rely 
on the re-summation of the S-wave projected chiral potential. Both are conceptually identical 
but differ in their treatment of the experimental data and fitting procedures. In our analysis, 
we have included the NLO models \IHWnlo{} from Ref.~\cite{Ikeda:2012au} 
and the models \GOi{} and \GOii{} from Ref.~\cite{Guo:2012vv}.
\item the Bonn approach \cite{Mai:2014xna} that does not rely on partial wave projection 
of the interaction kernel when solving the Bethe-Salpeter equation. In Ref.~\cite{Mai:2014xna}, 
two solutions (which we denote as \MMii{} and \MMiv{} here) of the global fits to the $K^{-}p$ experimental data 
were found compatible with the photoproduction data measured by the CLAS collaboration~\cite{Moriya:2013eb}. 
\item the Prague approach \cite{Cieply:2011nq} that differs from the other considered approaches 
by relying on effective separable meson-baryon potentials with off-shell form factors that also regularize 
the intermediate state Green function, which is equivalent to dimensional regularization used in the other 
approaches. In our analysis we use the NLO model \CSnlo{}, originally denoted as NLO30 in Ref.~\cite{Cieply:2011nq}. 
\end{itemize}

We refer the reader to the original papers \cite{Ikeda:2012au}, \cite{Guo:2012vv}, \cite{Mai:2014xna} 
and \cite{Cieply:2011nq} for the specific details of the considered approaches. Without dwelling upon  
any technicalities we proceed with a presentation of our findings in the next section and conclude 
the paper with a brief summary.

%
\section{Results and discussion}
\label{sec:res}

\begin{figure}[ht]
\centering
\resizebox{0.4\textwidth}{!}
{\includegraphics{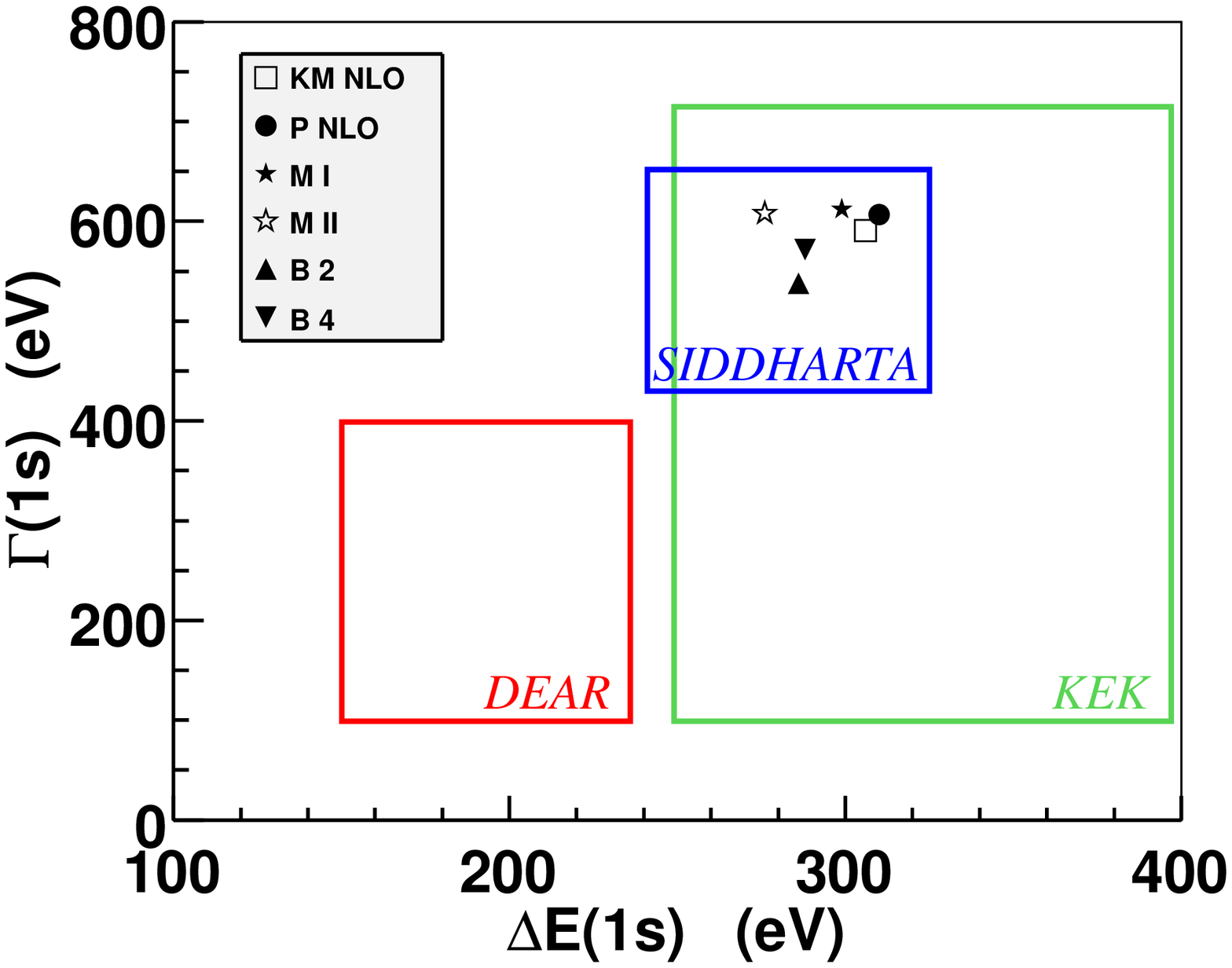}
}
\resizebox{0.4\textwidth}{!}
{\includegraphics{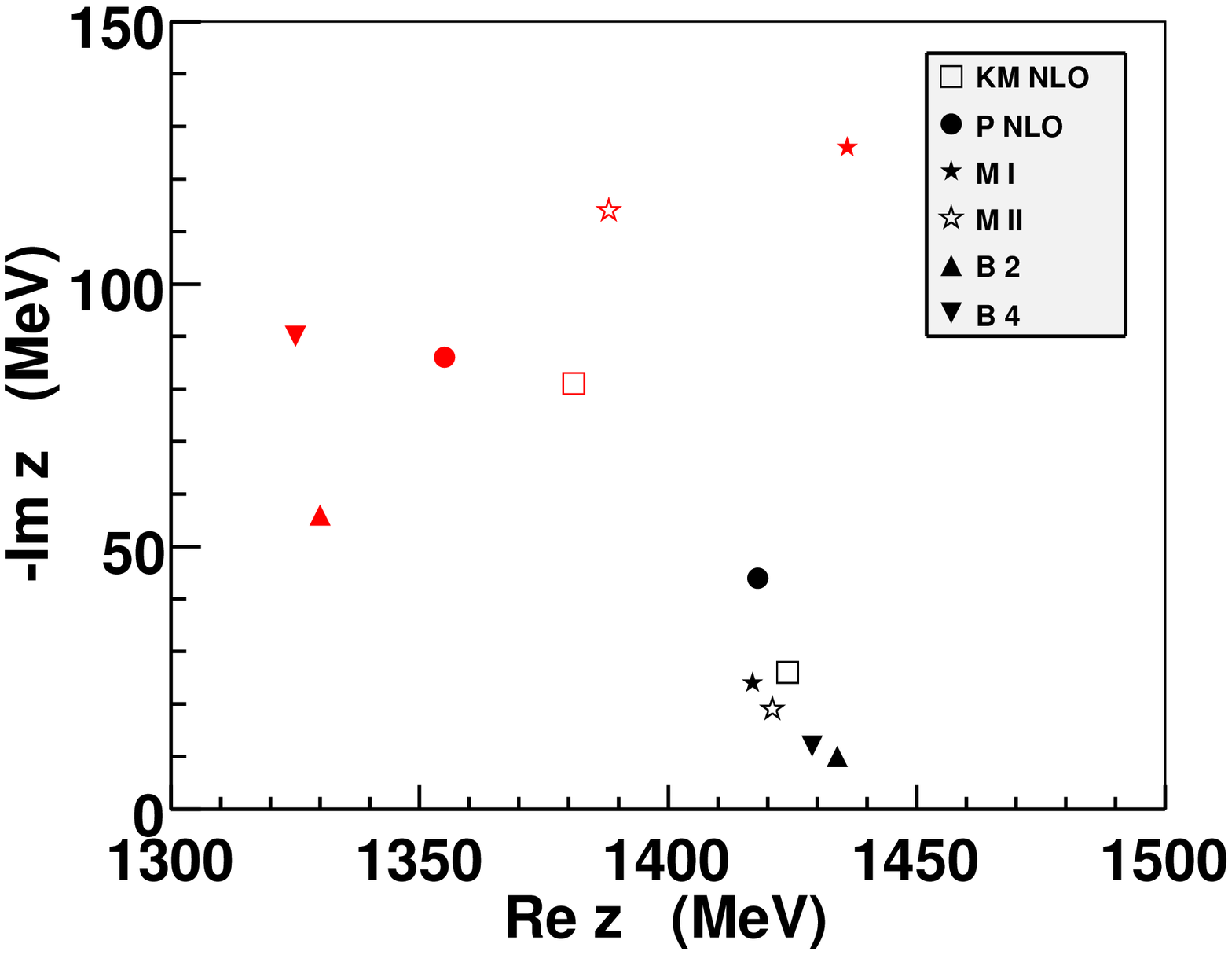}
} \\
\caption{Kaonic hydrogen characteristics (left panel) and positions of the poles assigned 
to $\Lambda(1405)$ (right panel) as generated by various theoretical approaches.}
\label{fig:poles}       
\end{figure}

The considered models represent the current state of theory on low energy meson-baryon interactions 
in the $S=-1$ sector and describe the $K^{-}p$ reactions data about equally well. This is demonstrated 
in the left panel of Fig.~\ref{fig:poles} which shows the theoretical predictions for the $1s$ level 
characteristics of the kaonic hydrogen, the energy shift $\Delta E(1s)$ and the absorption width 
$\Gamma (1s)$, both caused by the strong interaction. The rectangular areas drawn in the figure 
visualize the experimental progress with the rectangular boxes covering areas within one standard 
deviation of the experimental data taken from the KEK \cite{Iwasaki:1997wf}, DEAR \cite{Beer:2005qi} 
and SIDDHARTA \cite{Bazzi:2011zj} measurements. The theoretical approaches reproduce 
the most recent SIDDHARTA data quite well and are in very close agreement among each other. 
However, the same cannot be said concerning the positions of the poles assigned 
to the $\Lambda(1405)$ resonance and shown in the right panel of Fig.~\ref{fig:poles}. 
All models based on the chiral SU(3) dynamics generate invariably two poles 
in the $\pi\Sigma$-$\bar{K}N$ coupled channels sector. The models agree on the real part 
of the complex energy for the pole that couples more strongly to the $\bar{K}N$ channel 
and is generated at a higher energy of about 1420~MeV. Though, the imaginary part of the pole 
energy is not established so well and the position of the second pole varies from one model to another, 
apparently not constrained much by the experimental data. It was already shown in Ref.~\cite{Mai:2014xna}
that the new CLAS data on $\pi\Sigma$ photoproduction off proton \cite{Moriya:2013eb} provide additional 
constrains on the pole positions. On the other hand, the theoretical models still find it difficult 
to explain the peaks in the $\pi\Sigma$ mass spectra observed in the $pp$ collisions by the HADES 
experiment \cite{Agakishiev:2012xk}.

\begin{figure}[ht]
\centering
\resizebox{0.8\textwidth}{!}
{\includegraphics{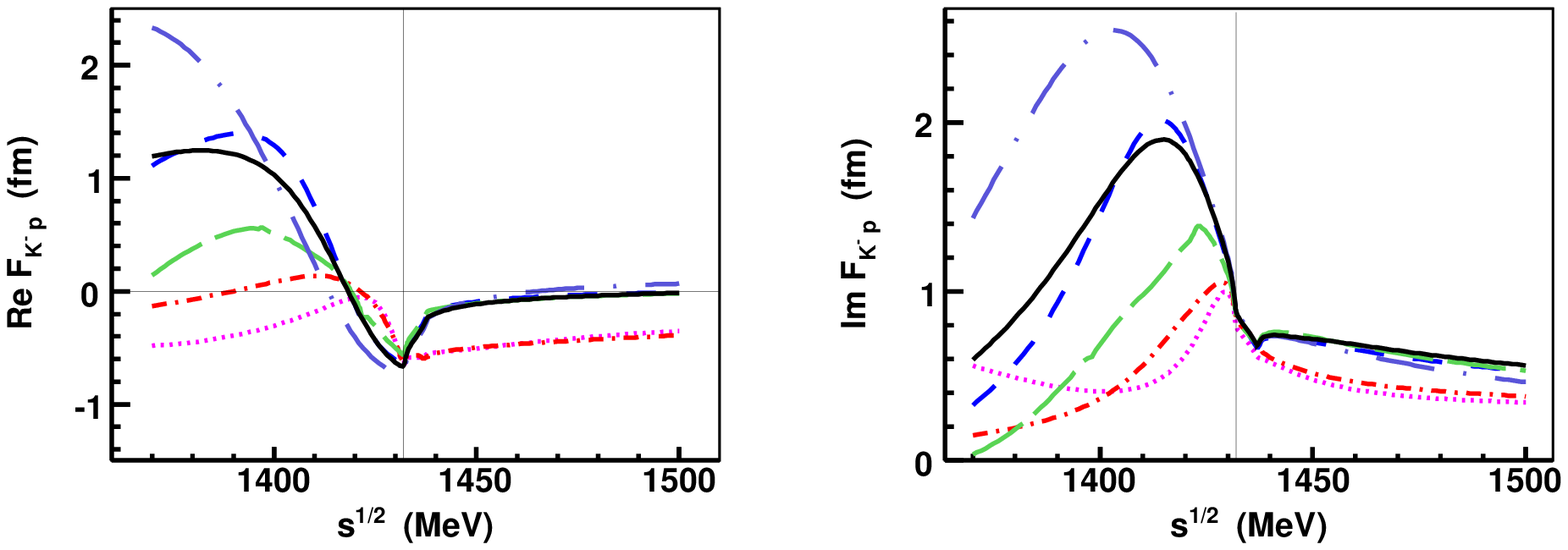}
}\\
\resizebox{0.8\textwidth}{!}
{\includegraphics{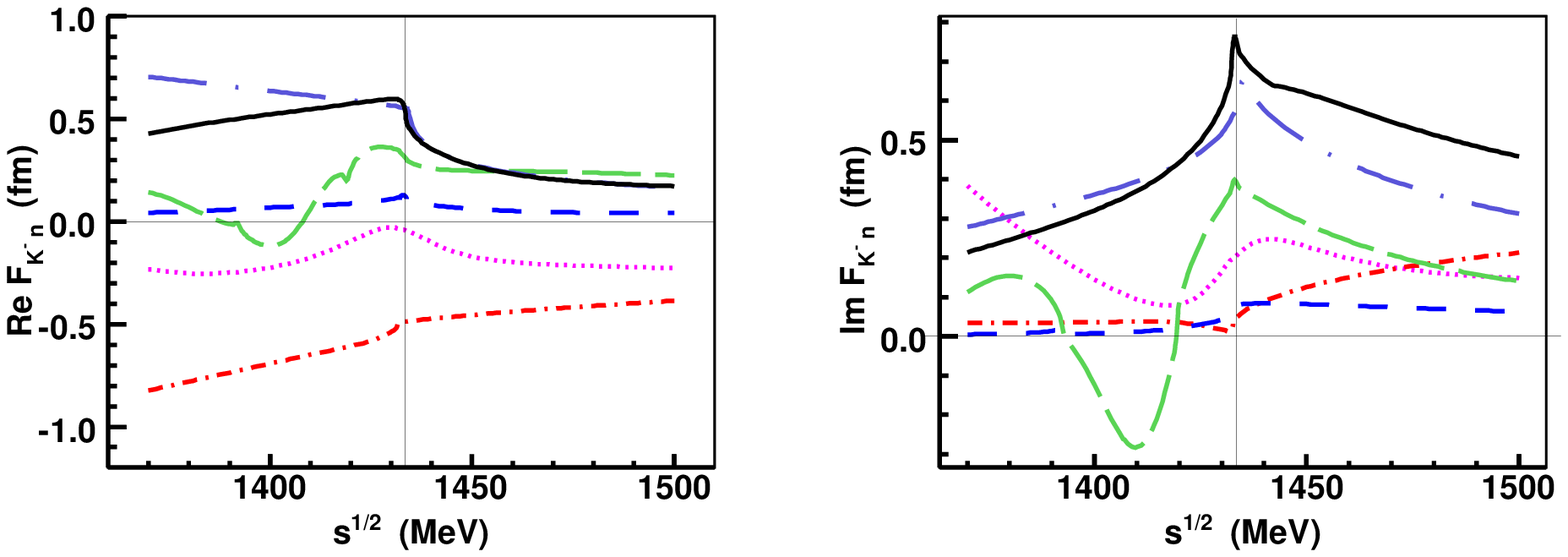}
}
\caption{The $K^{-}p$ (top panels) and $K^{-}n$ (bottom panels) elastic scattering amplitudes 
generated by the NLO approaches considered in our work. The various lines refer to the models:  
\MMii{} (dotted, purple), \MMiv{} (dot-dashed, red), \GOi{} (dashed, blue), 
\GOii{} (long-dashed, green), \CSnlo{} (dot-long-dashed, violet), \IHWnlo{} (continuous, black).
}
\label{fig:KNampl}       
\end{figure}

In Fig.~\ref{fig:KNampl} we also demonstrate that the considered approaches lead to very 
different predictions for the $K^{-}p$ amplitude extrapolated to sub-threshold 
energies as well as for the $K^{-}n$ amplitude. The theoretical ambiguities 
observed below the $\bar{K}N$ threshold are much larger then those standardly indicated 
by uncertainty bounds derived from variations of the $K^{-}p$ scattering length
within constraints enforced by the kaonic hydrogen data, see e.g.~Ref.~\cite{Ikeda:2012au}. 

Finally, in Ref.~\cite{Cieply:2016jby} we have also analyzed the origin of the poles 
of the scattering $T$-matrix generated by the theoretical models. There, we followed 
the pole movements to the so-called zero coupling limit (ZCL), in which the inter-channel couplings 
are switched off. Our findings are reviewed in the Table \ref{tab:ZCL} that shows the channels in which 
a pole assigned to a given resonance persists when the ZCL is reached. As there are two poles assigned 
to the $\Lambda(1405)$ we present them separately with the indexes $1$ and $2$. In the complex energy 
plane the $\Lambda_{1}(1405)$ pole is usually found at lower energy and further from the real axis 
than the $\Lambda_{2}(1405)$ pole. All models have the $\Lambda_{1}(1405)$ pole in the $\pi\Sigma$ channel 
when the inter-channel interactions are switched off. The $\Lambda_{2}(1405)$ pole couples most 
strongly to the $\bar{K}N$ channel, so it came as a surprise that the pole origin can be traced 
to the $\eta\Lambda$ channel in the ZCL for the \MMii{} and \GOii{} models. Thus, the other models 
that have the ZCL pole in the $\bar{K}N$ channel should be preferred if one anticipates 
a simplified picture of a $\bar{K}N$ bound state submerged in the $\pi\Sigma$ continuum 
\cite{Hyodo:2007jq}. We have also hinted in Ref.~\cite{Cieply:2016jby} at quite large 
NLO couplings occurring in the Bonn and Murcia models and being most likely responsible 
for an appearance of the $\eta\Lambda$ bound state in the ZCL.

In the isoscalar sector the models can also account for the $\Lambda(1670)$ resonance 
that emerges from the $K\Xi$ pole found in the ZCL. We have argued in Ref.~\cite{Cieply:2016jby} 
that an appearance of such pole is related to a particular condition imposed on a subtraction constant 
(or an inverse range in case of the Prague approach). If the condition is not met, the pole is missing 
as happens for the \IHWnlo{} and \MMii{} models. One should note, however, that with an exception  
of the Murcia approach the other approaches did not aim at describing the experimental data 
in the $\Lambda(1670)$ energy region, so it is not surprising that the pole is either completely 
missing or not at an appropriate position in those models.

\begin{table}[t!]
\centering
\caption{The origins (channels) of the poles generated by the considered models in which the poles 
are found when inter-channel couplings are switched off.}
\begin{tabular}{cc|cccccc}
&\multicolumn{7}{c}{\hspace{2cm}Models} \\
&                     &   \CSnlo{}   &   \IHWnlo{}  &   \GOi{}      &  \GOii{}      &  \MMii{}      &  \MMiv{}    \\
\cline{2-8}
\multirow{5}{*}{\rotatebox[origin=c]{90}{Resonances}}&$\Lambda_{1}(1405)$  &  $\pi\Sigma$ &  $\pi\Sigma$ &  $\pi\Sigma$  & $\pi\Sigma$   & $\pi\Sigma$   & $\pi\Sigma$ \\
&$\Lambda_{2}(1405)$  &  $\bar{K}N$  &  $\bar{K}N$  &  $\bar{K}N$   & $\eta\Lambda$ & $\eta\Lambda$ & $\bar{K}N$  \\
&$\Lambda(1670)$      &  $K\Xi$      &     ---      &  $K\Xi$       & $K\Xi$        &     ---       & $K\Xi$      \\
&$\bar{K}N(I=1)$      &  $\bar{K}N$  & $\eta\Sigma$ &  $\bar{K}N$   & $\bar{K}N$    &     ---       &   ---       \\
&$\Sigma(1750)$       &  $K\Xi$      &     ---      &  $K\Xi$       & $K\Xi$        &     ---       & $K\Xi$      
\end{tabular}
\label{tab:ZCL}       
\end{table}

Similarly, in the isovector sector the models can provide a pole which can be related 
to the $\Sigma(1750)$ resonance and the origin of this pole can be traced to the $K\Xi$ virtual (or bound) 
state in the ZCL. Several of the discussed models also predict an isovector $\bar{K}N$ 
pole located below the $\bar{K}N$ threshold at the Riemann sheet which is physical in the $\pi\Sigma$ 
and unphysical in the $\bar{K}N$ channel (it would be the third Riemann sheet if only these 
two channels were coupled). This pole emerges from an isovector $\bar{K}N$ virtual 
state generated in the ZCL by the Prague and Murcia models, though the Kyoto-Munich model has 
it in the $\eta\Sigma$ channel. We note that an existence of this pole was already witnessed 
in Refs.~\cite{Oller:2000fj}, \cite{Jido:2003cb} and \cite{Cieply:2011fy}. It is understood 
that it relates to the cusp structure in the energy dependence of the elastic $K^{-}n$ amplitude 
obtained for both, the \CSnlo{} and the \IHWnlo{} models as seen in Fig.~\ref{fig:KNampl}.

%
\section{Summary}
\label{sec:summary}

In the present work different versions of the modern chiral unitary approaches were compared directly 
for the first time. Our main observations are as follows:
\begin{itemize}
\item 
We have demonstrated that the available theoretical models lead to very different 
predictions for the elastic $K^{-}p$ and $K^{-}n$ amplitudes at sub-threshold energies. 
\item The tracking of the poles to the ZCL provides us with new insights related to the appearance of poles 
in a given approach. The procedure also reveals different concepts of forming the $\Lambda(1405)$.
\item Several models predict an existence of an isovector pole close to the $\bar{K}N$ threshold.
\end{itemize}

\begin{acknowledgement}
The authors acknowledge a collaboration with U.-G.~Mei{\ss}ner and J.~Smejkal who contributed 
to Ref.~\cite{Cieply:2016jby} the current presentation is based on.
\end{acknowledgement}

%
%

\end{document}